\documentclass[aps,prl,reprint,superscriptaddress,amsfonts,amssymb,amsmath,a4paper]{revtex4-2}

\usepackage{siunitx}
\usepackage{dcolumn}
\usepackage{hyperref}
\usepackage{graphicx}
\usepackage{dcolumn}
\usepackage{bm}
\usepackage{multirow}
\usepackage[xcdraw]{xcolor}
\usepackage{xcolor}
\usepackage[english]{babel}
\usepackage{array}
\usepackage{titlesec}

\newcolumntype{C}{>{\centering\arraybackslash}p{3em}}
\newcolumntype{E}{>{\centering\arraybackslash}p{5em}}

\DeclareSIUnit\bar{bar}
\hypersetup{
  colorlinks   = true, 
  urlcolor     = white!30!black, 
  linkcolor    = white!30!black, 
  citecolor   = white!30!black 
}

\makeatletter
\renewcommand\paragraph{%
  \@startsection
    {paragraph}%
    {4}%
    {0mm}%
    {0.4\baselineskip}%
    {-1em}%
    {
    \selectfont\bf}%
}%

\makeatother

\newcommand\dotafter[1]{#1.}
\titlespacing{\subsection}{0pt}{*0.5}{*1}
\titleformat{\subsection}[runin]
  {\normalfont\bfseries}{\thesubsection.}{3pt}{\dotafter}

\newcolumntype{d}[1]{D{.}{.}{#1}}

\begin{document}

\title{Superconductivity beyond the Pauli limit in  
 high-pressure CeSb$_2$}


\author{Oliver P.\ Squire}
\affiliation{Cavendish Laboratory, University of Cambridge, Cambridge CB3 0HE, UK}

\author{Stephen A.\ Hodgson}
\affiliation{Cavendish Laboratory, University of Cambridge, Cambridge CB3 0HE, UK}

\author{Jiasheng Chen}
\affiliation{Cavendish Laboratory, University of Cambridge, Cambridge CB3 0HE, UK}

\author{Vitaly Fedoseev}
\altaffiliation{Now at Dept. of Physics, Massachusetts Institute of Technology, Cambridge, MA 02139, USA}
\affiliation {Cavendish Laboratory, University of Cambridge, Cambridge CB3 0HE, UK}

\author{Christian K.\ de Podesta}
\affiliation {Cavendish Laboratory, University of Cambridge, Cambridge CB3 0HE, UK}

\author{Theodore I. Weinberger}
\affiliation {Cavendish Laboratory, University of Cambridge, Cambridge CB3 0HE, UK}

\author{Patricia L.\ Alireza}
\affiliation{Cavendish Laboratory, University of Cambridge, Cambridge CB3 0HE, UK}

\author{F.\ Malte Grosche}
\email[]{fmg12@cam.ac.uk}
\affiliation{Cavendish Laboratory, University of Cambridge, Cambridge CB3 0HE, UK}

\date{\today}

\begin{abstract}
\noindent 
We report the discovery of superconductivity at a pressure-induced magnetic quantum critical point in the Kondo-lattice system CeSb$_2$, sustained up to magnetic fields that exceed the conventional Pauli limit eight-fold. Like CeRh$_2$As$_2$, CeSb$_2$ is locally non-centrosymmetric around the Ce-site, but the evolution of critical fields and normal state properties as CeSb$_2$ is tuned through the quantum critical point motivates a fundamentally different  explanation for its resilience to applied field.

\end{abstract}

\maketitle
\noindent 
In an increasing number of materials  -- notably the new unconventional superconductors CeRh$_2$As$_2$ \cite{khim21} and UTe$_2$ \cite{ran19,aoki19} --  superconductivity is surprisingly resilient to magnetic field, and the temperature dependence of the upper critical field shows a rich and unexpected structure. 
This is important not just for applications in which high magnetic fields are required but also because the field resilience suggests that the superconducting Cooper pairs form triplet states, which may be exploited for quantum computing. In CeRh$_2$As$_2$, the postulated high field triplet state has been linked to a structural peculiarity, namely the lack of inversion symmetry around the crucially important Ce atoms, which underpin the electronic structure and the superconducting pairing mechanism. 


In the related, clean Kondo lattice material CeSb$_2$, we here report the discovery of superconductivity 
over a narrow pressure range that envelops a magnetic quantum critical point (qcp).  CeSb$_2$ displays a complex magnetic phase diagram with at least four magnetic phases and a ferromagnetic ground state \cite{canfield91,budko98,zhang17,liu20,trainer21}, all of which are initially robust under pressure, but its electronic and magnetic properties change profoundly  \cite{kagayama00,kagayama05} at the high pressures  considered here.
Like CeRh$_2$As$_2$, high pressure CeSb$_2$ lacks inversion symmetry around the Ce sites, and its upper critical field is strongly enhanced over expectations from elementary
theory. In contrast to CeRh$_2$As$_2$, however, signatures of a singlet-triplet transition under applied field are not observed in CeSb$_2$, suggesting that the critical field is instead boosted by a more general mechanism intrinsic to strong-coupling superconductivity involving ultra-heavy quasiparticles.


\begin{figure}
    \centering
    \includegraphics[width=\columnwidth]{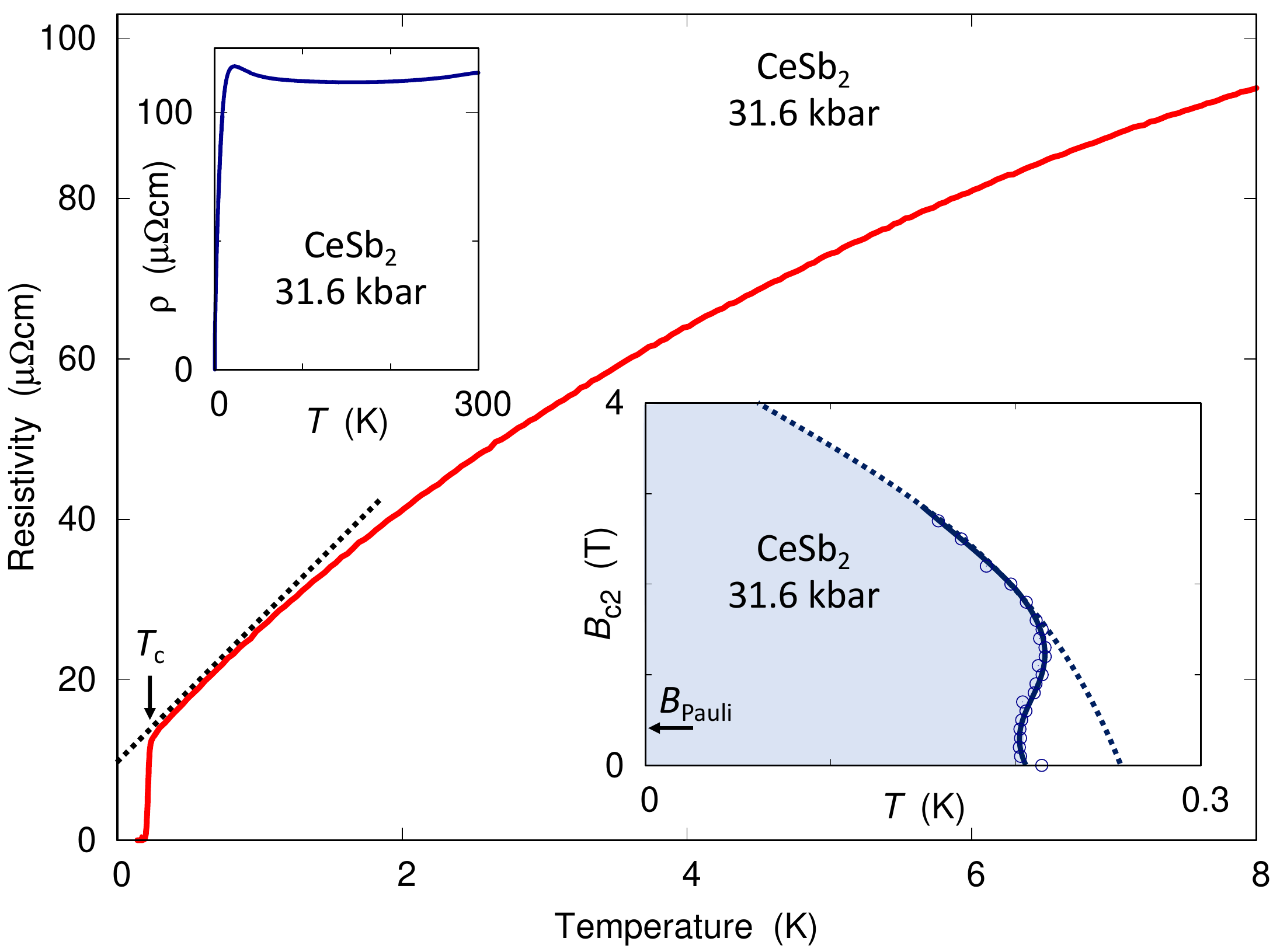}
    \caption{Superconductivity and anomalous normal state in high pressure CeSb$_2$. The variation of the resistivity $\rho$ with temperature $T$ shows negative curvature all the way down to a sharp transition to $\rho=0$ at $T_c \simeq \SI{0.22}{\kelvin}$. (left inset) $\rho(T)$ rises sharply to a shoulder at $\sim \SI{10}{\kelvin}$, reaches a shallow maximum at $\SI{22.5}{\kelvin}$ and then saturates, following a form typical for a Kondo lattice with a low effective bandwidth. (right inset) The resistive upper critical field follows an inverted `S'-shape at low fields and at intermediate fields takes on a large negative slope, which would extend to higher $T_c$ (dashed line) without the `S' anomaly. It far exceeds the Pauli paramagnetic limit $B_\text{Pauli} \simeq \SI{1.84}{\tesla \per \kelvin} ~T_c(B=0)$ (horizontal arrow).}
    \label{fig:Supercon}
\end{figure}

\begin{figure*}[t]
    \centering
    \includegraphics[width=\textwidth]{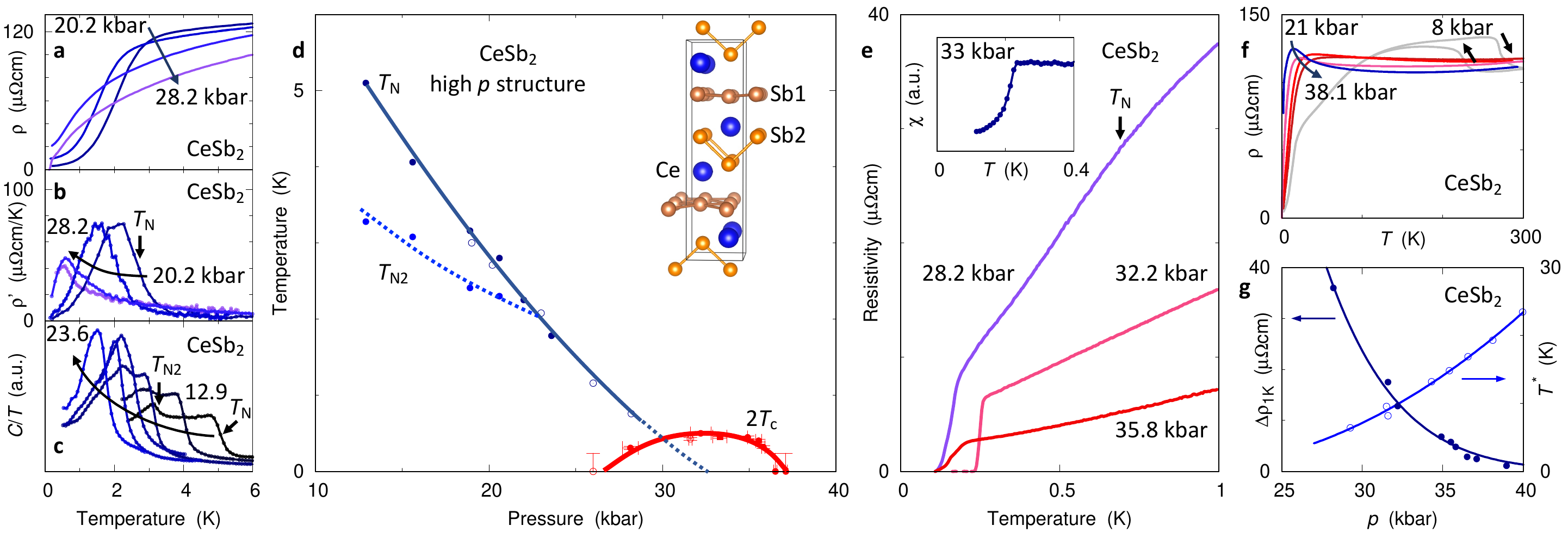}
    \caption{Pressure dependence of magnetic and superconducting states in high pressure CeSb$_2$. (a-c) Transition anomalies: (a) kink in $\rho(T)$, (b) associated jump in $\rho'=d\rho/dT$, (c) jumps in the heat capacity Sommerfeld ratio $C/T$. The $\rho(T)$ and $\rho'(T)$ data covers pressures from $\SI{20.2}{\kilo\bar}$ to $\SI{28.2}{\kilo\bar}$, above which these transition anomalies were no longer resolved. The heat capacity was measured at pressures ranging from $\SI{12.9}{\kilo\bar}$ to $\SI{23.6}{\kilo\bar}$. It shows a second transition anomaly at a lower temperature $T_{N2}$, in line with $\mu$SR data \cite{depodesta22}, which indicates that the low-$T$ state is magnetically ordered. (d) High pressure phase diagram of CeSb$_2$, showing the gradual suppression of the two magnetic transitions (full circles: from $C(T)$, empty circles: from $\rho(T)$) and a superconducting dome (full/empty symbols from $\rho(T)$ in two different samples, square: from magnetic susceptibility $\chi$ as in the inset to panel e). (inset to d) High pressure structure of CeSb$_2$. (e) $\rho(T)$ at different pressures straddling the qcp, showing the peak $T_c$ at $p_c \simeq \SI{32}{\kilo\bar}$, where a magnetic transition (arrow in $\SI{28.2}{\kilo\bar}$ data) extrapolates to zero, the quasilinear form of $\rho(T)$ at $p_c$, and the rapid suppression of $\rho(T)$ at low $T$ with increasing pressure. (inset to e) High-$p$ susceptibility data showing the superconducting transition. (f) Normal state resistivity up to room temperature, showing the hysteretic signature of the high $T$ structural transition at $\SI{8}{\kilo\bar}$ (arrows for cooling/warming data) and the very different form of $\rho(T)$ at higher pressures, typical for a Kondo lattice with a low characteristic temperature $T^*$. We estimate $T^*$ from the shoulder in $\rho(T)$, at which $\rho(T)$ reaches  $80\%$ of  $\text {max}(\rho)$. (g) Pressure dependence of $T^*$ and of the resistivity increment $\Delta\rho_\text{1K} = \rho(\text{1K})-\rho_0$, showing the rapid reduction of the $T$-dependence of $\rho(T)$ at low $T$ and the concomitant increase of $T^*$ with $p$.} 
    \label{fig:QCP}
\end{figure*}

\subsection{Methods}
High quality crystals of CeSb$_2$ with residual resistivity ratios $\mathrm{RRR} =\rho_{300}/\rho_0 \simeq 100$ were grown using standard self-flux techniques \cite{budko98} and characterised by powder x-ray diffraction, resistivity, magnetisation and heat capacity measurements. Piston-cylinder pressure cell measurements up to about \SI{28}{\kilo\bar} were carried out in a compound BeCu/MP35 cell \cite{walker99} with the superconducting transition temperature of Sn as the pressure gauge \cite{smith69}, whereas a wider pressure range up to \SI{40}{\kilo\bar} was accessed in moissanite anvil cells using  room temperature ruby fluorescence to determine the pressure. Glycerol was used as the pressure medium in both types of pressure cell. The crystal orientation reported in magnetic field studies ($c$-axis vs. in-plane) refers to the low pressure structure. 
The electrical resistivity was determined using a standard 4-terminal AC technique with a $\SI{3}{\micro\ampere}$ current at the lowest temperatures, and the magnetic susceptibility was measured using a mutual inductance technique with a pickup microcoil inside the high pressure sample volume \cite{alireza03}. The heat capacity was obtained from a $3\omega$ temperature modulation technique, oscillating the current in a thick film metal heater and in a Cernox thermometer closely connected to the sample at a frequency $\omega$, and picking up the third harmonic $\omega_3=3 \omega$ of the resulting thermometer signal \cite{SuppMat}. 
Measurements in a QD PPMS in the range \SI{2}{\kelvin}-\SI{300}{\kelvin} were complemented by low temperature studies in a cryogen-free ADR system (Dryogenic Measurement System, DMS) to $<\SI{0.1}{\kelvin}$ and in fields of up to $\SI{6}{\tesla}$.

%


\subsection{Superconductivity and anomalous normal state} 
The normal state in-plane resistivity in CeSb$_2$ at an applied pressure $p\simeq \SI{31.6}{\kilo\bar}$ displays a distinctly non-Fermi liquid, sub-linear temperature dependence $\rho(T)$ (\autoref{fig:Supercon}). The resistivity rises steeply at low $T$ and reaches a shallow maximum at $\SI{22.5}{\kelvin}$, above which it stays roughly constant up to room temperature (left inset in \autoref{fig:Supercon}), following a form familiar from other Ce or Yb-based Kondo lattice materials such as CeCu$_2$Si$_2$, CeCoIn$_5$, and YbRh$_2$Si$_2$ \cite{bellarbi84,petrovic01,malinowski05,trovarelli00}. It approaches saturation well below $\SI{10}{\kelvin}$, reaching $80\%$ of the maximum resistivity at $T^* \simeq\SI{8.2}{\kelvin}$. These temperatures are similar to those recorded in CeCu$_2$Si$_2$, CeCoIn$_5$ and YbRh$_2$Si$_2$, suggesting extremely strong electronic correlations, narrow renormalised bands and high quasiparticle masses in high-pressure CeSb$_2$. 

A sharp resistive transition with mid-point $T_c \simeq \SI{0.22}{\kelvin}$ (main plot in \autoref{fig:Supercon}) indicates superconductivity at very low temperatures, in line with the low electronic energy scales suggested by the normal state $\rho(T)$. 
Superconductivity proves surprisingly robust to applied magnetic  fields along the crystallographic $c$ direction (right inset in \autoref{fig:Supercon}). It persists to $>\SI{3}{\tesla}$ at low $T$, exceeding the Pauli paramagnetic limiting field, which is conventionally written as $B_\text{Pauli} =  \SI{1.84}{\tesla\per\kelvin} T_{\mathrm c} $ \cite{clogston62a,chandrasekhar62}, by nearly an order of magnitude. The in-plane upper critical field is similarly enhanced \cite{SuppMat}.

For small applied fields, $T_c$ is initially reduced, then rises again to a value slightly higher than the zero-field $T_c$, for $B\simeq \SI{1.5}{\tesla}$.
This produces an unusual, inverted `S'-shaped structure in the $B_{c2}(T)$ curve. The  inverted `S' structure  is observed at several other pressures $\le\SI{32.2}{\kilo\bar}$ but vanishes at higher pressures (see below). The sign reversal of $dB_{c2}/dT$, which is $>0$  over an intermediate field range, points towards an underlying, field tuned phase transition within the normal state \cite{SuppMat}.

\begin{figure}
    \centering
    \includegraphics[width=\columnwidth]{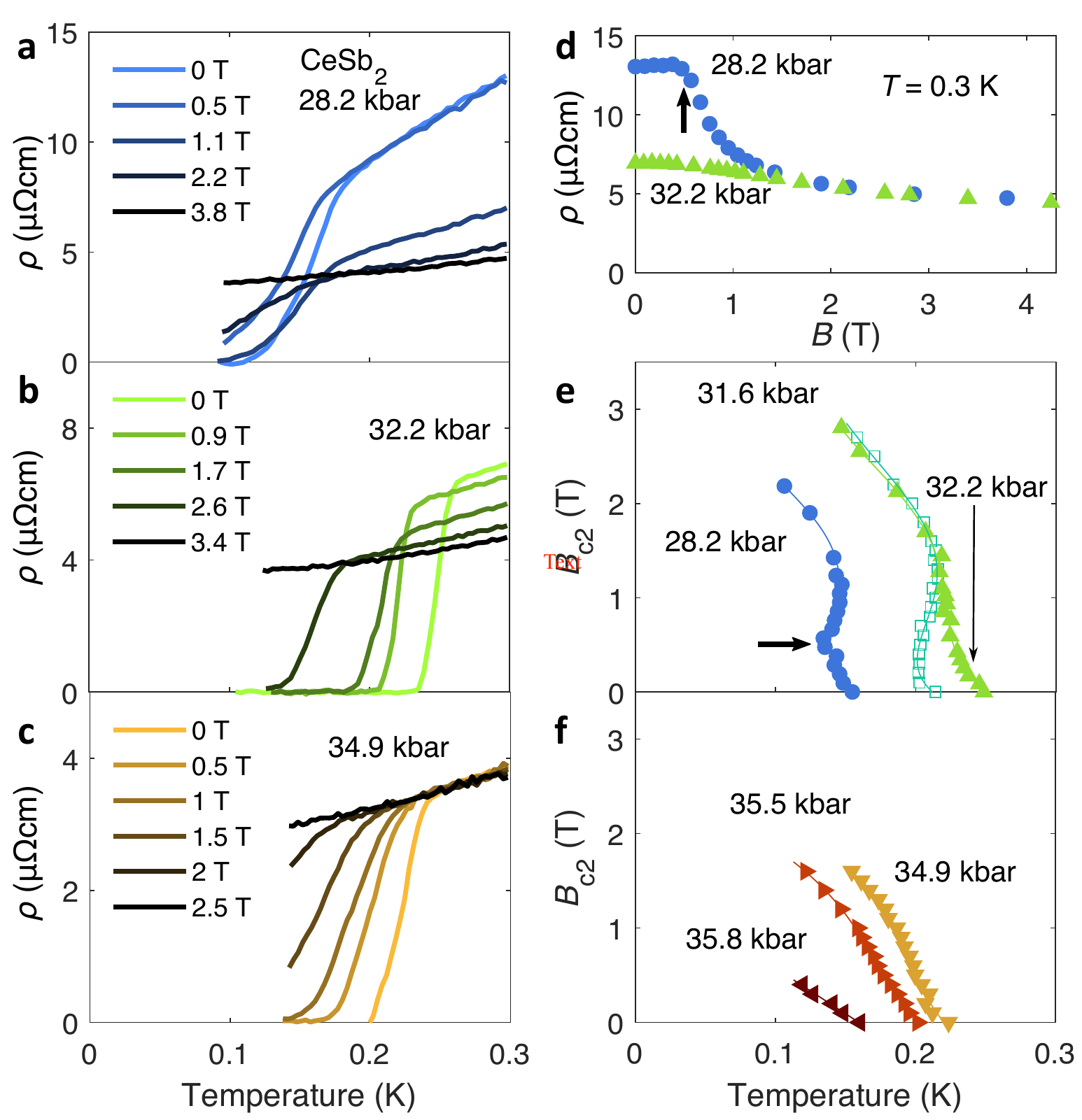}
    \caption{(a-c) Response of the superconducting transition in CeSb$_2$ to magnetic field applied along the $c$-axis, for $p<p_c$ (a), for $p\simeq p_c$ (b) and for $p>p_c$ (c). 
    (d) The magnetoresistance at $T>T_c$ displays a distinct kink at $\simeq \SI{0.5}{\tesla}$ (vertical arrow) for $p<p_c$, indicating a field-induced transition out of the magnetically ordered state. (e)  $B_{c2}(T)$ curves extracted from the mid-point of the resistive transition display a pronounced inverted `S'-shape with a local minimum of $T_c$ at $\simeq \SI{0.5}{\tesla}$ for $p=\SI{28.2}{\kilo\bar}$ (horizontal arrow), which corresponds to the kink field in panel (d). (f) At $p>p_c$ the $B_{c2}(T)$ curves revert to a more conventional form. }
    \label{fig:CritField}
\end{figure}	

\subsection{Quantum critical point}
 Distinct transition anomalies are indeed observed at pressures less than $p_c \simeq \SI{32}{\kilo\bar}$ (\autoref{fig:QCP}a-c).
Electric transport measurements for $p<p_c$ find a kink in $\rho(T)$ at low $T$, which causes a jump in the $T$-derivative of the resistivity $\rho'(T)$ (\autoref{fig:QCP}a-b). Heat capacity measurements under pressure likewise display a jump in $C(T)$ (\autoref{fig:QCP}c) at a transition temperature $T_N$ that is consistent with that of the kink in $\rho(T)$. Heat capacity data furthermore show evidence for a weaker, second transition at a lower temperature $T_{N2}$, which merges with $T_N$ as pressure is increased. 
Both step-like heat capacity signatures suggest second-order transitions. High pressure muon spin rotation studies indicate two distinct magnetically ordered states associated with $T_{N}$ and $T_{N2}$ \cite{depodesta22}. The detection of two magnetic transitions is reminiscent of the CeCu$_2$(Si/Ge)$_2$ system \cite{yuan03} and of YbRh$_2$Si$_2$ under pressure \cite{trovarelli00}. 
The magnetic transition signatures extrapolate to zero temperature at a magnetic quantum critical point near $\SI{32}{\kilo\bar}$. 
The superconducting transition has likewise been tracked in high pressure transport measurements using two anvil cells and a susceptibility measurement in a third anvil cell (inset of \autoref{fig:QCP}e).
Following the magnetic and superconducting transition signatures as functions of pressure results in the phase diagram (\autoref{fig:QCP}d), which shows a superconducting dome tightly confined to the immediate vicinity of a magnetic quantum critical point, indicating a prominent role for magnetic fluctuations in the superconducting pairing mechanism.  

Normal and superconducting properties of CeSb$_2$ evolve rapidly with pressure (\autoref{fig:QCP}e-g). The low $T$ resistivity takes a quasi-linear $T$ dependence near $p_c$ (\autoref{fig:QCP}e), which saturates to a nearly constant resistivity (\autoref{fig:QCP}f) above  a low $T^*\sim \SI{10}{\kelvin}$. 
The low $T$ slope of $\rho(T)$, measured by the resistivity increment $\Delta\rho_\text{1K}=\rho(\SI{1}{\kelvin})-\rho_0$ over the extrapolated residual resistivity $\rho_0$ diminishes rapidly with increasing pressure. This is accompanied by a steep increase in  $T^*$, demonstrating that compression under applied pressure strongly increases the effective electronic bandwidth in CeSb$_2$ (\autoref{fig:QCP}g).

\subsection{High pressure structure} 
CeSb$_2$ 
forms in the orthorhombic SmSb$_2$ structure (space group 64), which lacks inversion symmetry around the Ce site but  is centrosymmetric around the center of the unit cell. 
Transport measurements at intermediate pressures $\SI{6}{\kilo\bar}< p <\SI{17}{\kilo\bar}$ show a highly hysteretic resistivity anomaly (e.g. $\SI{8}{\kilo\bar}$ data in \autoref{fig:QCP}f), which shifts to lower temperature with increasing pressure \cite{kagayama00} and disappears beyond $\SI{17}{\kilo\bar}$, where the low-$T$ state differs profoundly from the low-$T$ state at ambient pressure \cite{kagayama05}. High pressure X-ray diffraction \cite{depodesta22} has established that this anomaly signals a first-order structural phase transition, which at low $T$ 
is complete by about $\SI{17}{\kilo\bar}$. 
The superconducting and magnetic states discussed above are therefore all associated with the high pressure structure of CeSb$_2$. 
The rare earth (R) diantimonides RSb$_2$ adopt a variety of structure types, all of which lack inversion symmetry around the rare earth site: SmSb$_2$ (like CeSb$_2$ at $p=0$), HoSb$_2$ (orthorhombic, space-group 21), EuSb$_2$ (monoclinic, space-group 11) and YbSb$_2$ (orthorhombic, space-group 63).  The X-ray data and  {\em ab initio} DFT calculations in \cite{depodesta22} unambiguously rule out the SmSb$_2$ and HoSb$_2$ structures for high pressure CeSb$_2$ and favour the YbSb$_2$ structure (inset in \autoref{fig:QCP}d). 
%

\subsection{Critical fields} 
The locally non-centroysmmetric structure of high pressure CeSb$_2$ invites comparison to CeRh$_2$As$_2$ \cite{khim21,mockli21a,landaeta22,cavanagh22,hafner22,kibune22} and  other unconventional superconductors such as UTe$_2$, UGe$_2$ and UPt$_3$ (e.g. \cite{hazra22}) when considering the response  to applied magnetic field.
Both the form of the critical field curve $B_{c2}(T)$ in CeSb$_2$ and the magnitude of the upper critical field are unusual. We consider first the inverted `S'-shaped form for $B_{c2}(T)$ displayed in the inset of \autoref{fig:Supercon}. 
The initial reduction, then increase of $T_c$ with field is most pronounced at the lowest pressure at which full resistive transitions could be observed ($\SI{28.2}{\kilo\bar}$, \autoref{fig:CritField}a). It is already weaker at $\SI{31.6}{\kilo\bar}$ (\autoref{fig:Supercon}) and weaker still close to the qcp, at $\SI{32.2}{\kilo\bar}$ (\autoref{fig:CritField}b, e).
Comparing $B_{c2}(T)$ at these last two pressures (\autoref{fig:CritField}e) shows that near the qcp, the critical field curves converge on a single line at high fields but differ at low fields. 
At pressures above $p_c$, the critical field curves gradually change into the conventional form (\autoref{fig:CritField}f). 
The relative reduction of $T_c$ at low fields $< \SI{0.5}{\tesla}$ for $p<p_c$  could be seen as a signature of a field-induced transition between two distinct superconducting states, as in CeRh$_2$As$_2$ \cite{khim21}, 
or it might result from a field-induced magnetic transition. 
The step-like magnetoresistance anomaly at $\SI{28.2}{\kilo\bar}$ shown in \autoref{fig:CritField}d points towards the second scenario. The transition field of $\simeq \SI{0.5}{\tesla}$ (vertical arrow) corresponds to the minimum $T_c$ in the $\SI{28.2}{\kilo\bar}$ critical field curve in \autoref{fig:CritField}e (horizontal arrow). 
These findings suggest that  the inverted `S' shape of $B_{c2}(T)$   on the ordered side of the qcp results from the interplay between applied field and the magnetic spin fluctuation spectrum: tuning the system  out of the magnetically ordered state with increasing field enhances order parameter fluctuations and the associated pairing interaction, thereby strengthening superconductivity.
A similar explanation has been advanced in pressurised UGe$_2$ \cite{sheikin01}.
%

\begin{table}[tbp]
  \centering
  \caption{Critical field data for selected heavy fermion superconductors. $T_c$, initial slope of the upper critical field $B_{c2}'$, experimental $B_{c2}$ in the low-$T$ limit and $C/T$ at $T_c$ have been extracted from the literature. The Pauli limit $B_\text{Pauli}$ is calculated as $\SI{1.84}{\tesla\per\kelvin}~ T_c$. Elementary theory predicts that $\sqrt{B_{c2}'/T_c} \propto C/T$ \cite{SuppMat}, as is indeed roughly confirmed by the tabulated data. Applying this analysis to CeSb$_2$ near $p_c$, at $\SI{34.9}{\kilo\bar}$, produces an estimate for $C/T$ of $\sim \SI{1.2}{\joule\per \mol\kelvin^2}$.}
\setlength{\tabcolsep}{2pt}
\begin{tabular}{| l | c | c | c | c | c | c |}
\hline
 & $T_c$  & $B_{c2}' $ & $B_\text{Pauli} $ & $B_{c2}(0)$ & $\sqrt{B_{c2}' /T_c}$ & $C/T  $ \\
 & K & $\SI{}{\tesla/\kelvin}$ & $\SI{}{\tesla}$ & $\SI{}{\tesla}$ & $\SI{}{\tesla^{1/2}\per\kelvin}$ & $\SI{}{\joule\per\mole\kelvin^2}$ \\
 \hline
    CeCoIn$_5$ \cite{miclea06} & 2.2 & 30.5    & 4.05  & 11.5  & 3.7   & 0.3   \\
    CeCu$_2$Si$_2$ \cite{kittaka16} & 0.6  & 35     & 1.10  & 1.9   & 7.6   & 0.7    \\
    CeRh$_2$As$_2$ \cite{khim21}  & 0.26 & 97     & 0.48  & 14    & 19.3  & 2   \\
    CeSb$_2$ ($\simeq p_c$) & 0.22 & 30    & 0.40  & $>3$   & 11.8  &   1.2 (est.)    \\
    UPt$_3$ \cite{chen84} & 0.52 & 6.3     & 0.96  & 1.8   & 3.5   & 0.4  \\
    UBe$_{13}$ \cite{thomas96} & 0.95 & 45     & 1.75  & 14    & 6.9   & 1    \\
 \hline
    \end{tabular}%
  \label{tab:HFSCs}%
\end{table}%

Considering next the eight-fold enhancement of $B_{c2}$ over the conventional Pauli limit $B_\text{Pauli} =  \SI{1.84}{\tesla\per\kelvin}~T_c$ \cite{clogston62,chandrasekhar62} in CeSb$_2$, we note that
moderate violations of the Pauli limit are common in Ce-based heavy fermion materials such as CeCoIn$_5$ and CeCu$_2$Si$_2$ (\autoref{tab:HFSCs}) 
without necessarily being taken as evidence for triplet pairing. 
The ratio of the high initial slope $B_{c2}'$ over $T_{c}$ in compressed CeSb$_2$ indicates a very high Sommerfeld ratio $C/T \sim \SI{1.2}{\joule/\mole\kelvin^2}$ (\autoref{tab:HFSCs})  \cite{SuppMat}. It is larger than the corresponding ratios in UPt$_3$, CeCoIn$_5$, CeCu$_2$Si$_2$, and UBe$_{13}$, suggesting that the quasiparticles underlying superconductivity in high pressure CeSb$_2$ are among the heaviest ever recorded in a superconducting heavy fermion material. This is significant, because 
theoretical studies 
 \cite{schossmann89,carbotte90} indicate that violations of Pauli limiting may generally be expected in superconductors with large mass renormalisation,  irrespective of whether the pairing is mediated by phonons or spin-fluctuations and whether the pairing state has $s$-wave or $d$-wave symmetry \cite{perez-gonzalez96}. 
The original calculation of the conventional Pauli limiting field  \cite{clogston62,chandrasekhar62} balances the superconducting condensation energy against the magnetic energy involved in changing the spin alignment of the paired electrons in an applied field. The former depends on the energy gap, the latter on the spin susceptibility. Although some uncertainty in the latter arises from imprecise knowledge of the conduction electron $g$-factor, this would have to be $<< 1$ to explain substantially enhanced Pauli limiting fields, which  is difficult to justify: strong anisotropy of the $g$-factor is ruled out by the large $B_{c2}$ observed for $B \perp c$ \cite{SuppMat}. In strong-coupling superconductors the balance between condensation energy and magnetic energy needs to be modified both on the side of the condensation energy, because the energy gap may be far larger than the BCS relation $\Delta = 1.76 k_B T_c$ suggests, and on the side of the magnetic energy, because the spin susceptibility is reduced below the Pauli susceptibility indicated by the quasiparticle density of states by as much as the interaction-induced mass enhancement. In model calculations, this causes the Pauli limit to be boosted to about $1.5\SI{}{\tesla\per\kelvin} T_c m^*/m_b$, with $m^*$ the renormalised quasiparticle mass and $m_b$ the bare band mass \cite{schossmann89}. In UBe$_{13}$, the eight-fold enhancement of $B_{c2}$ over the conventional Pauli limit  (\autoref{tab:HFSCs}) has been interpreted likewise \cite{thomas96} in terms of a strong-coupling calculation, and a similar boost to the limiting field was found in a calculation for spin-fluctuation induced $d$-wave pairing \cite{perez-gonzalez96}. In this approach, resilience to high fields is achieved by gradually admixing a frequency-odd triplet pairing state into the underlying frequency-even singlet pairing state  \cite{schossmann86,matsumoto12} (see also \cite{aperis15,aperis20} for material-specific calculations). This general route contrasts starkly with the scenario advanced for CeRh$_2$As$_2$ (e.g. \cite{khim21}), which is predicated on its locally non-centrosymmetric structure. 
In heavy fermion materials such as CeSb$_2$, a quantitative calculation is hindered by the similar magnitudes of the Zeeman energy  at $B_{c2}$ and electronic as well as bosonic energy scales, by the effect of the applied field on the pairing interaction, by the highly anomalous normal state, which deviates profoundly from expectations of Fermi liquid theory, and by our incomplete understanding of the origins of mass renormalization and pairing interaction, which do not align completely. The intriguing suggestion that 
increasing admixture of odd-frequency, triplet superconductivity can boost the critical field in strongly correlated materials 
should be tested in more detailed theoretical and computational investigations. 
High-pressure CeSb$_2$ emerges as a clean, ultra-heavy fermion system with superconductivity forming out of a pronounced non-Fermi liquid state and an upper critical field far beyond the Pauli limit. Because the qcp underlying the superconducting dome can in CeSb$_2$ be  crossed under pressure, this material supplies an excellent test case for refining our understanding of unconventional superconductivity. 
Our findings suggest that strong mass renormalization boosts the magnitude of $B_{c2}$ without the need for a singlet-triplet phase transition as reported in CeRh$_2$As$_2$, and that the interplay between applied field, magnetic order and the associated magnetic fluctuations can explain the evolution of $B_{c2}(T)$ across the qcp.




\begin{acknowledgments}
We thank, in particular, A. Chubukov and G. Lonzarich for helpful discussions and Z. Feng and Y. Zou for early work on the structural and magnetic transitions in CeSb$_2$. The project was supported by the EPSRC of the UK (grants no. EP/K012894 and EP/P023290/1) and by Trinity College.
\end{acknowledgments}

\bibliography{CeSb2-Paper}

\begin{thebibliography}{46}%
\makeatletter
\providecommand \@ifxundefined [1]{%
 \@ifx{#1\undefined}
}%
\providecommand \@ifnum [1]{%
 \ifnum #1\expandafter \@firstoftwo
 \else \expandafter \@secondoftwo
 \fi
}%
\providecommand \@ifx [1]{%
 \ifx #1\expandafter \@firstoftwo
 \else \expandafter \@secondoftwo
 \fi
}%
\providecommand \natexlab [1]{#1}%
\providecommand \enquote  [1]{``#1''}%
\providecommand \bibnamefont  [1]{#1}%
\providecommand \bibfnamefont [1]{#1}%
\providecommand \citenamefont [1]{#1}%
\providecommand \href@noop [0]{\@secondoftwo}%
\providecommand \href [0]{\begingroup \@sanitize@url \@href}%
\providecommand \@href[1]{\@@startlink{#1}\@@href}%
\providecommand \@@href[1]{\endgroup#1\@@endlink}%
\providecommand \@sanitize@url [0]{\catcode `\\12\catcode `\$12\catcode
  `\&12\catcode `\#12\catcode `\^12\catcode `\_12\catcode `\%12\relax}%
\providecommand \@@startlink[1]{}%
\providecommand \@@endlink[0]{}%
\providecommand \url  [0]{\begingroup\@sanitize@url \@url }%
\providecommand \@url [1]{\endgroup\@href {#1}{\urlprefix }}%
\providecommand \urlprefix  [0]{URL }%
\providecommand \Eprint [0]{\href }%
\providecommand \doibase [0]{https://doi.org/}%
\providecommand \selectlanguage [0]{\@gobble}%
\providecommand \bibinfo  [0]{\@secondoftwo}%
\providecommand \bibfield  [0]{\@secondoftwo}%
\providecommand \translation [1]{[#1]}%
\providecommand \BibitemOpen [0]{}%
\providecommand \bibitemStop [0]{}%
\providecommand \bibitemNoStop [0]{.\EOS\space}%
\providecommand \EOS [0]{\spacefactor3000\relax}%
\providecommand \BibitemShut  [1]{\csname bibitem#1\endcsname}%
\let\auto@bib@innerbib\@empty
\bibitem [{\citenamefont {Khim}\ \emph {et~al.}(2021)\citenamefont {Khim},
  \citenamefont {Landaeta}, \citenamefont {Banda}, \citenamefont {Bannor},
  \citenamefont {Brando}, \citenamefont {Brydon}, \citenamefont {Hafner},
  \citenamefont {Küchler}, \citenamefont {Cardoso-Gil}, \citenamefont
  {Stockert}, \citenamefont {Mackenzie}, \citenamefont {Agterberg},
  \citenamefont {Geibel},\ and\ \citenamefont {Hassinger}}]{khim21}%
  \BibitemOpen
  \bibfield  {author} {\bibinfo {author} {\bibfnamefont {S.}~\bibnamefont
  {Khim}}, \bibinfo {author} {\bibfnamefont {J.}~\bibnamefont {Landaeta}},
  \bibinfo {author} {\bibfnamefont {J.}~\bibnamefont {Banda}}, \bibinfo
  {author} {\bibfnamefont {N.}~\bibnamefont {Bannor}}, \bibinfo {author}
  {\bibfnamefont {M.}~\bibnamefont {Brando}}, \bibinfo {author} {\bibfnamefont
  {P.}~\bibnamefont {Brydon}}, \bibinfo {author} {\bibfnamefont
  {D.}~\bibnamefont {Hafner}}, \bibinfo {author} {\bibfnamefont
  {R.}~\bibnamefont {Küchler}}, \bibinfo {author} {\bibfnamefont
  {R.}~\bibnamefont {Cardoso-Gil}}, \bibinfo {author} {\bibfnamefont
  {U.}~\bibnamefont {Stockert}}, \bibinfo {author} {\bibfnamefont {A.~P.}\
  \bibnamefont {Mackenzie}}, \bibinfo {author} {\bibfnamefont {D.~F.}\
  \bibnamefont {Agterberg}}, \bibinfo {author} {\bibfnamefont {C.}~\bibnamefont
  {Geibel}},\ and\ \bibinfo {author} {\bibfnamefont {E.}~\bibnamefont
  {Hassinger}},\ }\href@noop {} {\bibfield  {journal} {\bibinfo  {journal}
  {Science}\ }\textbf {\bibinfo {volume} {373}},\ \bibinfo {pages} {1012}
  (\bibinfo {year} {2021})}\BibitemShut {NoStop}%
\bibitem [{\citenamefont {Ran}\ \emph {et~al.}(2019)\citenamefont {Ran},
  \citenamefont {Eckberg}, \citenamefont {Ding}, \citenamefont {Furukawa},
  \citenamefont {Metz}, \citenamefont {Saha}, \citenamefont {Liu},
  \citenamefont {Zic}, \citenamefont {Kim}, \citenamefont {Paglione},\ and\
  \citenamefont {Butch}}]{ran19}%
  \BibitemOpen
  \bibfield  {author} {\bibinfo {author} {\bibfnamefont {S.}~\bibnamefont
  {Ran}}, \bibinfo {author} {\bibfnamefont {C.}~\bibnamefont {Eckberg}},
  \bibinfo {author} {\bibfnamefont {Q.-P.}\ \bibnamefont {Ding}}, \bibinfo
  {author} {\bibfnamefont {Y.}~\bibnamefont {Furukawa}}, \bibinfo {author}
  {\bibfnamefont {T.}~\bibnamefont {Metz}}, \bibinfo {author} {\bibfnamefont
  {S.~R.}\ \bibnamefont {Saha}}, \bibinfo {author} {\bibfnamefont {I.-L.}\
  \bibnamefont {Liu}}, \bibinfo {author} {\bibfnamefont {M.}~\bibnamefont
  {Zic}}, \bibinfo {author} {\bibfnamefont {H.}~\bibnamefont {Kim}}, \bibinfo
  {author} {\bibfnamefont {J.}~\bibnamefont {Paglione}},\ and\ \bibinfo
  {author} {\bibfnamefont {N.~P.}\ \bibnamefont {Butch}},\ }\href
  {https://www.science.org/doi/10.1126/science.aav8645} {\bibfield  {journal}
  {\bibinfo  {journal} {Science}\ }\textbf {\bibinfo {volume} {365}},\ \bibinfo
  {pages} {684} (\bibinfo {year} {2019})}\BibitemShut {NoStop}%
\bibitem [{\citenamefont {Aoki}\ \emph {et~al.}(2019)\citenamefont {Aoki},
  \citenamefont {Nakamura}, \citenamefont {Honda}, \citenamefont {Li},
  \citenamefont {Homma}, \citenamefont {Shimizu}, \citenamefont {Sato},
  \citenamefont {Knebel}, \citenamefont {Brison}, \citenamefont {Pourret},
  \citenamefont {Braithwaite}, \citenamefont {Lapertot}, \citenamefont {Niu},
  \citenamefont {Vali{\v s}ka}, \citenamefont {Harima},\ and\ \citenamefont
  {Flouquet}}]{aoki19}%
  \BibitemOpen
  \bibfield  {author} {\bibinfo {author} {\bibfnamefont {D.}~\bibnamefont
  {Aoki}}, \bibinfo {author} {\bibfnamefont {A.}~\bibnamefont {Nakamura}},
  \bibinfo {author} {\bibfnamefont {F.}~\bibnamefont {Honda}}, \bibinfo
  {author} {\bibfnamefont {D.}~\bibnamefont {Li}}, \bibinfo {author}
  {\bibfnamefont {Y.}~\bibnamefont {Homma}}, \bibinfo {author} {\bibfnamefont
  {Y.}~\bibnamefont {Shimizu}}, \bibinfo {author} {\bibfnamefont {Y.~J.}\
  \bibnamefont {Sato}}, \bibinfo {author} {\bibfnamefont {G.}~\bibnamefont
  {Knebel}}, \bibinfo {author} {\bibfnamefont {J.-P.}\ \bibnamefont {Brison}},
  \bibinfo {author} {\bibfnamefont {A.}~\bibnamefont {Pourret}}, \bibinfo
  {author} {\bibfnamefont {D.}~\bibnamefont {Braithwaite}}, \bibinfo {author}
  {\bibfnamefont {G.}~\bibnamefont {Lapertot}}, \bibinfo {author}
  {\bibfnamefont {Q.}~\bibnamefont {Niu}}, \bibinfo {author} {\bibfnamefont
  {M.}~\bibnamefont {Vali{\v s}ka}}, \bibinfo {author} {\bibfnamefont
  {H.}~\bibnamefont {Harima}},\ and\ \bibinfo {author} {\bibfnamefont
  {J.}~\bibnamefont {Flouquet}},\ }\href
  {https://journals.jps.jp/doi/10.7566/JPSJ.88.043702} {\bibfield  {journal}
  {\bibinfo  {journal} {J. Phys. Soc. Jpn.}\ }\textbf {\bibinfo {volume}
  {88}},\ \bibinfo {pages} {043702} (\bibinfo {year} {2019})}\BibitemShut
  {NoStop}%
\bibitem [{\citenamefont {Canfield}\ \emph {et~al.}(1991)\citenamefont
  {Canfield}, \citenamefont {Thompson},\ and\ \citenamefont
  {Fisk}}]{canfield91}%
  \BibitemOpen
  \bibfield  {author} {\bibinfo {author} {\bibfnamefont {P.~C.}\ \bibnamefont
  {Canfield}}, \bibinfo {author} {\bibfnamefont {J.~D.}\ \bibnamefont
  {Thompson}},\ and\ \bibinfo {author} {\bibfnamefont {Z.}~\bibnamefont
  {Fisk}},\ }\href {http://aip.scitation.org/doi/10.1063/1.350071} {\bibfield
  {journal} {\bibinfo  {journal} {Journal of Applied Physics}\ }\textbf
  {\bibinfo {volume} {70}},\ \bibinfo {pages} {5992} (\bibinfo {year}
  {1991})}\BibitemShut {NoStop}%
\bibitem [{\citenamefont {Bud'ko}\ \emph {et~al.}(1998)\citenamefont {Bud'ko},
  \citenamefont {Canfield}, \citenamefont {Mielke},\ and\ \citenamefont
  {Lacerda}}]{budko98}%
  \BibitemOpen
  \bibfield  {author} {\bibinfo {author} {\bibfnamefont {S.~L.}\ \bibnamefont
  {Bud'ko}}, \bibinfo {author} {\bibfnamefont {P.~C.}\ \bibnamefont
  {Canfield}}, \bibinfo {author} {\bibfnamefont {C.~H.}\ \bibnamefont
  {Mielke}},\ and\ \bibinfo {author} {\bibfnamefont {A.~H.}\ \bibnamefont
  {Lacerda}},\ }\href {https://link.aps.org/doi/10.1103/PhysRevB.57.13624}
  {\bibfield  {journal} {\bibinfo  {journal} {Phys. Rev. B}\ }\textbf {\bibinfo
  {volume} {57}},\ \bibinfo {pages} {13624} (\bibinfo {year}
  {1998})}\BibitemShut {NoStop}%
\bibitem [{\citenamefont {Zhang}\ \emph {et~al.}(2017)\citenamefont {Zhang},
  \citenamefont {Zhu}, \citenamefont {Hu}, \citenamefont {Tan}, \citenamefont
  {Xie}, \citenamefont {Feng}, \citenamefont {Qin}, \citenamefont {Zhang},
  \citenamefont {Liu}, \citenamefont {Song}, \citenamefont {Luo}, \citenamefont
  {Zhang},\ and\ \citenamefont {Lai}}]{zhang17}%
  \BibitemOpen
  \bibfield  {author} {\bibinfo {author} {\bibfnamefont {Y.}~\bibnamefont
  {Zhang}}, \bibinfo {author} {\bibfnamefont {X.}~\bibnamefont {Zhu}}, \bibinfo
  {author} {\bibfnamefont {B.}~\bibnamefont {Hu}}, \bibinfo {author}
  {\bibfnamefont {S.}~\bibnamefont {Tan}}, \bibinfo {author} {\bibfnamefont
  {D.}~\bibnamefont {Xie}}, \bibinfo {author} {\bibfnamefont {W.}~\bibnamefont
  {Feng}}, \bibinfo {author} {\bibfnamefont {L.}~\bibnamefont {Qin}}, \bibinfo
  {author} {\bibfnamefont {W.}~\bibnamefont {Zhang}}, \bibinfo {author}
  {\bibfnamefont {Y.}~\bibnamefont {Liu}}, \bibinfo {author} {\bibfnamefont
  {H.}~\bibnamefont {Song}}, \bibinfo {author} {\bibfnamefont {L.}~\bibnamefont
  {Luo}}, \bibinfo {author} {\bibfnamefont {Z.}~\bibnamefont {Zhang}},\ and\
  \bibinfo {author} {\bibfnamefont {X.}~\bibnamefont {Lai}},\ }\href
  {https://iopscience.iop.org/article/10.1088/1674-1056/26/6/067102} {\bibfield
   {journal} {\bibinfo  {journal} {Chinese Phys. B}\ }\textbf {\bibinfo
  {volume} {26}},\ \bibinfo {pages} {067102} (\bibinfo {year}
  {2017})}\BibitemShut {NoStop}%
\bibitem [{\citenamefont {Liu}\ \emph {et~al.}(2020)\citenamefont {Liu},
  \citenamefont {Wang}, \citenamefont {Radelytskyi}, \citenamefont {Zhang},
  \citenamefont {Meven}, \citenamefont {Deng}, \citenamefont {Zhu},
  \citenamefont {Su}, \citenamefont {Zhu}, \citenamefont {Tan},\ and\
  \citenamefont {Schneidewind}}]{liu20}%
  \BibitemOpen
  \bibfield  {author} {\bibinfo {author} {\bibfnamefont {B.}~\bibnamefont
  {Liu}}, \bibinfo {author} {\bibfnamefont {L.}~\bibnamefont {Wang}}, \bibinfo
  {author} {\bibfnamefont {I.}~\bibnamefont {Radelytskyi}}, \bibinfo {author}
  {\bibfnamefont {Y.}~\bibnamefont {Zhang}}, \bibinfo {author} {\bibfnamefont
  {M.}~\bibnamefont {Meven}}, \bibinfo {author} {\bibfnamefont
  {H.}~\bibnamefont {Deng}}, \bibinfo {author} {\bibfnamefont {F.}~\bibnamefont
  {Zhu}}, \bibinfo {author} {\bibfnamefont {Y.}~\bibnamefont {Su}}, \bibinfo
  {author} {\bibfnamefont {X.}~\bibnamefont {Zhu}}, \bibinfo {author}
  {\bibfnamefont {S.}~\bibnamefont {Tan}},\ and\ \bibinfo {author}
  {\bibfnamefont {A.}~\bibnamefont {Schneidewind}},\ }\href
  {https://iopscience.iop.org/article/10.1088/1361-648X/ab9bcd} {\bibfield
  {journal} {\bibinfo  {journal} {J. Phys.: Condens. Matter}\ }\textbf
  {\bibinfo {volume} {32}},\ \bibinfo {pages} {405605} (\bibinfo {year}
  {2020})}\BibitemShut {NoStop}%
\bibitem [{\citenamefont {Trainer}\ \emph {et~al.}(2021)\citenamefont
  {Trainer}, \citenamefont {Abel}, \citenamefont {Bud'ko}, \citenamefont
  {Canfield},\ and\ \citenamefont {Wahl}}]{trainer21}%
  \BibitemOpen
  \bibfield  {author} {\bibinfo {author} {\bibfnamefont {C.}~\bibnamefont
  {Trainer}}, \bibinfo {author} {\bibfnamefont {C.}~\bibnamefont {Abel}},
  \bibinfo {author} {\bibfnamefont {S.~L.}\ \bibnamefont {Bud'ko}}, \bibinfo
  {author} {\bibfnamefont {P.~C.}\ \bibnamefont {Canfield}},\ and\ \bibinfo
  {author} {\bibfnamefont {P.}~\bibnamefont {Wahl}},\ }\href
  {https://link.aps.org/doi/10.1103/PhysRevB.104.205134} {\bibfield  {journal}
  {\bibinfo  {journal} {Phys. Rev. B}\ }\textbf {\bibinfo {volume} {104}},\
  \bibinfo {pages} {205134} (\bibinfo {year} {2021})}\BibitemShut {NoStop}%
\bibitem [{\citenamefont {Kagayama}\ \emph {et~al.}(2000)\citenamefont
  {Kagayama}, \citenamefont {Oomi}, \citenamefont {Bud'ko},\ and\ \citenamefont
  {Canfield}}]{kagayama00}%
  \BibitemOpen
  \bibfield  {author} {\bibinfo {author} {\bibfnamefont {T.}~\bibnamefont
  {Kagayama}}, \bibinfo {author} {\bibfnamefont {G.}~\bibnamefont {Oomi}},
  \bibinfo {author} {\bibfnamefont {S.}~\bibnamefont {Bud'ko}},\ and\ \bibinfo
  {author} {\bibfnamefont {P.}~\bibnamefont {Canfield}},\ }\href@noop {}
  {\bibfield  {journal} {\bibinfo  {journal} {Physica B: Condensed Matter}\
  }\textbf {\bibinfo {volume} {281}},\ \bibinfo {pages} {90} (\bibinfo {year}
  {2000})}\BibitemShut {NoStop}%
\bibitem [{\citenamefont {Kagayama}\ \emph {et~al.}(2005)\citenamefont
  {Kagayama}, \citenamefont {Uwatoko}, \citenamefont {Bud'ko},\ and\
  \citenamefont {Canfield}}]{kagayama05}%
  \BibitemOpen
  \bibfield  {author} {\bibinfo {author} {\bibfnamefont {T.}~\bibnamefont
  {Kagayama}}, \bibinfo {author} {\bibfnamefont {Y.}~\bibnamefont {Uwatoko}},
  \bibinfo {author} {\bibfnamefont {S.}~\bibnamefont {Bud'ko}},\ and\ \bibinfo
  {author} {\bibfnamefont {P.}~\bibnamefont {Canfield}},\ }\href
  {https://linkinghub.elsevier.com/retrieve/pii/S0921452605001250} {\bibfield
  {journal} {\bibinfo  {journal} {Physica B: Condensed Matter}\ }\textbf
  {\bibinfo {volume} {359--361}},\ \bibinfo {pages} {320} (\bibinfo {year}
  {2005})}\BibitemShut {NoStop}%
\bibitem [{\citenamefont {{de Podesta}}\ \emph {et~al.}(2022)\citenamefont {{de
  Podesta}}, \citenamefont {Weinberger}, \citenamefont {Squire}, \citenamefont
  {Feng}, \citenamefont {Chen}, \citenamefont {Lampronti}, \citenamefont
  {Khasanov},\ and\ \citenamefont {Grosche}}]{depodesta22}%
  \BibitemOpen
  \bibfield  {author} {\bibinfo {author} {\bibfnamefont {C.}~\bibnamefont {{de
  Podesta}}}, \bibinfo {author} {\bibfnamefont {T.}~\bibnamefont {Weinberger}},
  \bibinfo {author} {\bibfnamefont {O.}~\bibnamefont {Squire}}, \bibinfo
  {author} {\bibfnamefont {Z.}~\bibnamefont {Feng}}, \bibinfo {author}
  {\bibfnamefont {J.}~\bibnamefont {Chen}}, \bibinfo {author} {\bibfnamefont
  {G.~I.}\ \bibnamefont {Lampronti}}, \bibinfo {author} {\bibfnamefont
  {R.}~\bibnamefont {Khasanov}},\ and\ \bibinfo {author} {\bibfnamefont
  {F.~M.}\ \bibnamefont {Grosche}},\ }\href@noop {} {\bibfield  {journal}
  {\bibinfo  {journal} {to be publ.}\ } (\bibinfo {year} {2022})}\BibitemShut
  {NoStop}%
\bibitem [{\citenamefont {Walker}(1999)}]{walker99}%
  \BibitemOpen
  \bibfield  {author} {\bibinfo {author} {\bibfnamefont {I.~R.}\ \bibnamefont
  {Walker}},\ }\href@noop {} {\bibfield  {journal} {\bibinfo  {journal} {Rev.
  Sci. Instrum.}\ }\textbf {\bibinfo {volume} {70}},\ \bibinfo {pages} {3402}
  (\bibinfo {year} {1999})}\BibitemShut {NoStop}%
\bibitem [{\citenamefont {Smith}\ \emph {et~al.}(1969)\citenamefont {Smith},
  \citenamefont {Chu},\ and\ \citenamefont {Maple}}]{smith69}%
  \BibitemOpen
  \bibfield  {author} {\bibinfo {author} {\bibfnamefont {T.}~\bibnamefont
  {Smith}}, \bibinfo {author} {\bibfnamefont {C.}~\bibnamefont {Chu}},\ and\
  \bibinfo {author} {\bibfnamefont {M.}~\bibnamefont {Maple}},\ }\href
  {https://linkinghub.elsevier.com/retrieve/pii/0011227569902604} {\bibfield
  {journal} {\bibinfo  {journal} {Cryogenics}\ }\textbf {\bibinfo {volume}
  {9}},\ \bibinfo {pages} {53} (\bibinfo {year} {1969})}\BibitemShut {NoStop}%
\bibitem [{\citenamefont {Alireza}\ and\ \citenamefont
  {Julian}(2003)}]{alireza03}%
  \BibitemOpen
  \bibfield  {author} {\bibinfo {author} {\bibfnamefont {P.~L.}\ \bibnamefont
  {Alireza}}\ and\ \bibinfo {author} {\bibfnamefont {S.~R.}\ \bibnamefont
  {Julian}},\ }\href {http://aip.scitation.org/doi/10.1063/1.1614861}
  {\bibfield  {journal} {\bibinfo  {journal} {Rev. Sci. Instr.}\ }\textbf
  {\bibinfo {volume} {74}},\ \bibinfo {pages} {4728} (\bibinfo {year}
  {2003})}\BibitemShut {NoStop}%
\bibitem [{Sup(2022)}]{SuppMat}%
  \BibitemOpen
  \href@noop {} {\bibfield  {journal} {\bibinfo  {journal} {{See Supplementary
  Material at [URL], which includes Refs.
  \cite{helfand66,tinkham04,orlando79,sullivan68,gati19} for further details of
  the in-plane critical field, estimates based on the critical field data,
  implications of the sign-reversal of $B_{c2}'(T)$, and the temperature
  modulation heat capacity method}}\ } (\bibinfo {year} {2022})}\BibitemShut
  {NoStop}%
\bibitem [{\citenamefont {Bellarbi}\ \emph {et~al.}(1984)\citenamefont
  {Bellarbi}, \citenamefont {Benoit}, \citenamefont {Jaccard}, \citenamefont
  {Mignot},\ and\ \citenamefont {Braun}}]{bellarbi84}%
  \BibitemOpen
  \bibfield  {author} {\bibinfo {author} {\bibfnamefont {B.}~\bibnamefont
  {Bellarbi}}, \bibinfo {author} {\bibfnamefont {A.}~\bibnamefont {Benoit}},
  \bibinfo {author} {\bibfnamefont {D.}~\bibnamefont {Jaccard}}, \bibinfo
  {author} {\bibfnamefont {J.~M.}\ \bibnamefont {Mignot}},\ and\ \bibinfo
  {author} {\bibfnamefont {H.~F.}\ \bibnamefont {Braun}},\ }\href
  {https://link.aps.org/doi/10.1103/PhysRevB.30.1182} {\bibfield  {journal}
  {\bibinfo  {journal} {Phys. Rev. B}\ }\textbf {\bibinfo {volume} {30}},\
  \bibinfo {pages} {1182} (\bibinfo {year} {1984})}\BibitemShut {NoStop}%
\bibitem [{\citenamefont {Petrovic}\ \emph {et~al.}(2001)\citenamefont
  {Petrovic}, \citenamefont {Pagliuso}, \citenamefont {Hundley}, \citenamefont
  {Movshovich}, \citenamefont {Sarrao}, \citenamefont {Thompson}, \citenamefont
  {Fisk},\ and\ \citenamefont {Monthoux}}]{petrovic01}%
  \BibitemOpen
  \bibfield  {author} {\bibinfo {author} {\bibfnamefont {C.}~\bibnamefont
  {Petrovic}}, \bibinfo {author} {\bibfnamefont {P.~G.}\ \bibnamefont
  {Pagliuso}}, \bibinfo {author} {\bibfnamefont {M.~F.}\ \bibnamefont
  {Hundley}}, \bibinfo {author} {\bibfnamefont {R.}~\bibnamefont {Movshovich}},
  \bibinfo {author} {\bibfnamefont {J.~L.}\ \bibnamefont {Sarrao}}, \bibinfo
  {author} {\bibfnamefont {J.~D.}\ \bibnamefont {Thompson}}, \bibinfo {author}
  {\bibfnamefont {Z.}~\bibnamefont {Fisk}},\ and\ \bibinfo {author}
  {\bibfnamefont {P.}~\bibnamefont {Monthoux}},\ }\href
  {http://stacks.iop.org/0953-8984/13/i=17/a=103?key=crossref.5adfc86f1b51e9b9c9a7a974642037ec}
  {\bibfield  {journal} {\bibinfo  {journal} {J. Phys.: Condens. Matter}\
  }\textbf {\bibinfo {volume} {13}},\ \bibinfo {pages} {L337} (\bibinfo {year}
  {2001})}\BibitemShut {NoStop}%
\bibitem [{\citenamefont {Malinowski}\ \emph {et~al.}(2005)\citenamefont
  {Malinowski}, \citenamefont {Hundley}, \citenamefont {Capan}, \citenamefont
  {Ronning}, \citenamefont {Movshovich}, \citenamefont {Moreno}, \citenamefont
  {Sarrao},\ and\ \citenamefont {Thompson}}]{malinowski05}%
  \BibitemOpen
  \bibfield  {author} {\bibinfo {author} {\bibfnamefont {A.}~\bibnamefont
  {Malinowski}}, \bibinfo {author} {\bibfnamefont {M.~F.}\ \bibnamefont
  {Hundley}}, \bibinfo {author} {\bibfnamefont {C.}~\bibnamefont {Capan}},
  \bibinfo {author} {\bibfnamefont {F.}~\bibnamefont {Ronning}}, \bibinfo
  {author} {\bibfnamefont {R.}~\bibnamefont {Movshovich}}, \bibinfo {author}
  {\bibfnamefont {N.~O.}\ \bibnamefont {Moreno}}, \bibinfo {author}
  {\bibfnamefont {J.~L.}\ \bibnamefont {Sarrao}},\ and\ \bibinfo {author}
  {\bibfnamefont {J.~D.}\ \bibnamefont {Thompson}},\ }\href
  {https://link.aps.org/doi/10.1103/PhysRevB.72.184506} {\bibfield  {journal}
  {\bibinfo  {journal} {Phys. Rev. B}\ }\textbf {\bibinfo {volume} {72}},\
  \bibinfo {pages} {184506} (\bibinfo {year} {2005})}\BibitemShut {NoStop}%
\bibitem [{\citenamefont {Trovarelli}\ \emph {et~al.}(2000)\citenamefont
  {Trovarelli}, \citenamefont {Geibel}, \citenamefont {Mederle}, \citenamefont
  {Langhammer}, \citenamefont {Grosche}, \citenamefont {Gegenwart},
  \citenamefont {Lang}, \citenamefont {Sparn},\ and\ \citenamefont
  {Steglich}}]{trovarelli00}%
  \BibitemOpen
  \bibfield  {author} {\bibinfo {author} {\bibfnamefont {O.}~\bibnamefont
  {Trovarelli}}, \bibinfo {author} {\bibfnamefont {C.}~\bibnamefont {Geibel}},
  \bibinfo {author} {\bibfnamefont {S.}~\bibnamefont {Mederle}}, \bibinfo
  {author} {\bibfnamefont {C.}~\bibnamefont {Langhammer}}, \bibinfo {author}
  {\bibfnamefont {F.~M.}\ \bibnamefont {Grosche}}, \bibinfo {author}
  {\bibfnamefont {P.}~\bibnamefont {Gegenwart}}, \bibinfo {author}
  {\bibfnamefont {M.}~\bibnamefont {Lang}}, \bibinfo {author} {\bibfnamefont
  {G.}~\bibnamefont {Sparn}},\ and\ \bibinfo {author} {\bibfnamefont
  {F.}~\bibnamefont {Steglich}},\ }\href
  {https://link.aps.org/doi/10.1103/PhysRevLett.85.626} {\bibfield  {journal}
  {\bibinfo  {journal} {Phys. Rev. Lett.}\ }\textbf {\bibinfo {volume} {85}},\
  \bibinfo {pages} {626} (\bibinfo {year} {2000})}\BibitemShut {NoStop}%
\bibitem [{\citenamefont {Clogston}(1962)}]{clogston62a}%
  \BibitemOpen
  \bibfield  {author} {\bibinfo {author} {\bibfnamefont {A.~M.}\ \bibnamefont
  {Clogston}},\ }\href {https://link.aps.org/doi/10.1103/PhysRevLett.9.266}
  {\bibfield  {journal} {\bibinfo  {journal} {Phys. Rev. Lett.}\ }\textbf
  {\bibinfo {volume} {9}},\ \bibinfo {pages} {266} (\bibinfo {year}
  {1962})}\BibitemShut {NoStop}%
\bibitem [{\citenamefont {Chandrasekhar}(1962)}]{chandrasekhar62}%
  \BibitemOpen
  \bibfield  {author} {\bibinfo {author} {\bibfnamefont {B.~S.}\ \bibnamefont
  {Chandrasekhar}},\ }\href {http://aip.scitation.org/doi/10.1063/1.1777362}
  {\bibfield  {journal} {\bibinfo  {journal} {Appl. Phys. Lett.}\ }\textbf
  {\bibinfo {volume} {1}},\ \bibinfo {pages} {7} (\bibinfo {year}
  {1962})}\BibitemShut {NoStop}%
\bibitem [{\citenamefont {Yuan}\ \emph {et~al.}(2003)\citenamefont {Yuan},
  \citenamefont {Grosche}, \citenamefont {Deppe}, \citenamefont {Geibel},
  \citenamefont {Sparn},\ and\ \citenamefont {Steglich}}]{yuan03}%
  \BibitemOpen
  \bibfield  {author} {\bibinfo {author} {\bibfnamefont {H.}~\bibnamefont
  {Yuan}}, \bibinfo {author} {\bibfnamefont {F.}~\bibnamefont {Grosche}},
  \bibinfo {author} {\bibfnamefont {M.}~\bibnamefont {Deppe}}, \bibinfo
  {author} {\bibfnamefont {C.}~\bibnamefont {Geibel}}, \bibinfo {author}
  {\bibfnamefont {G.}~\bibnamefont {Sparn}},\ and\ \bibinfo {author}
  {\bibfnamefont {F.}~\bibnamefont {Steglich}},\ }\href@noop {} {\bibfield
  {journal} {\bibinfo  {journal} {Science}\ }\textbf {\bibinfo {volume}
  {302}},\ \bibinfo {pages} {2104} (\bibinfo {year} {2003})}\BibitemShut
  {NoStop}%
\bibitem [{\citenamefont {Möckli}\ and\ \citenamefont
  {Ramires}(2021)}]{mockli21a}%
  \BibitemOpen
  \bibfield  {author} {\bibinfo {author} {\bibfnamefont {D.}~\bibnamefont
  {Möckli}}\ and\ \bibinfo {author} {\bibfnamefont {A.}~\bibnamefont
  {Ramires}},\ }\href@noop {} {\bibfield  {journal} {\bibinfo  {journal}
  {Physical Review B}\ }\textbf {\bibinfo {volume} {104}},\ \bibinfo {pages}
  {134517} (\bibinfo {year} {2021})}\BibitemShut {NoStop}%
\bibitem [{\citenamefont {Landaeta}\ \emph {et~al.}(2022)\citenamefont
  {Landaeta}, \citenamefont {Khanenko}, \citenamefont {Cavanagh}, \citenamefont
  {Geibel}, \citenamefont {Khim}, \citenamefont {Mishra}, \citenamefont
  {Sheikin}, \citenamefont {Brydon}, \citenamefont {Agterberg}, \citenamefont
  {Brando},\ and\ \citenamefont {Hassinger}}]{landaeta22}%
  \BibitemOpen
  \bibfield  {author} {\bibinfo {author} {\bibfnamefont {J.~F.}\ \bibnamefont
  {Landaeta}}, \bibinfo {author} {\bibfnamefont {P.}~\bibnamefont {Khanenko}},
  \bibinfo {author} {\bibfnamefont {D.~C.}\ \bibnamefont {Cavanagh}}, \bibinfo
  {author} {\bibfnamefont {C.}~\bibnamefont {Geibel}}, \bibinfo {author}
  {\bibfnamefont {S.}~\bibnamefont {Khim}}, \bibinfo {author} {\bibfnamefont
  {S.}~\bibnamefont {Mishra}}, \bibinfo {author} {\bibfnamefont
  {I.}~\bibnamefont {Sheikin}}, \bibinfo {author} {\bibfnamefont {P.~M.~R.}\
  \bibnamefont {Brydon}}, \bibinfo {author} {\bibfnamefont {D.~F.}\
  \bibnamefont {Agterberg}}, \bibinfo {author} {\bibfnamefont {M.}~\bibnamefont
  {Brando}},\ and\ \bibinfo {author} {\bibfnamefont {E.}~\bibnamefont
  {Hassinger}},\ }\href {http://arxiv.org/abs/2204.07975} {\bibfield  {journal}
  {\bibinfo  {journal} {ArXiv220407975 Cond-Mat}\ } (\bibinfo {year}
  {2022})}\BibitemShut {NoStop}%
\bibitem [{\citenamefont {Cavanagh}\ \emph {et~al.}(2022)\citenamefont
  {Cavanagh}, \citenamefont {Shishidou}, \citenamefont {Weinert}, \citenamefont
  {Brydon},\ and\ \citenamefont {Agterberg}}]{cavanagh22}%
  \BibitemOpen
  \bibfield  {author} {\bibinfo {author} {\bibfnamefont {D.~C.}\ \bibnamefont
  {Cavanagh}}, \bibinfo {author} {\bibfnamefont {T.}~\bibnamefont {Shishidou}},
  \bibinfo {author} {\bibfnamefont {M.}~\bibnamefont {Weinert}}, \bibinfo
  {author} {\bibfnamefont {P.~M.~R.}\ \bibnamefont {Brydon}},\ and\ \bibinfo
  {author} {\bibfnamefont {D.~F.}\ \bibnamefont {Agterberg}},\ }\href@noop {}
  {\bibfield  {journal} {\bibinfo  {journal} {Phys. Rev. B}\ }\textbf {\bibinfo
  {volume} {105}},\ \bibinfo {pages} {L020505} (\bibinfo {year}
  {2022})}\BibitemShut {NoStop}%
\bibitem [{\citenamefont {Hafner}\ \emph {et~al.}(2022)\citenamefont {Hafner},
  \citenamefont {Khanenko}, \citenamefont {Eljaouhari}, \citenamefont
  {Küchler}, \citenamefont {Banda}, \citenamefont {Bannor}, \citenamefont
  {Lühmann}, \citenamefont {Landaeta}, \citenamefont {Mishra}, \citenamefont
  {Sheikin}, \citenamefont {Hassinger}, \citenamefont {Khim}, \citenamefont
  {Geibel}, \citenamefont {Zwicknagl},\ and\ \citenamefont
  {Brando}}]{hafner22}%
  \BibitemOpen
  \bibfield  {author} {\bibinfo {author} {\bibfnamefont {D.}~\bibnamefont
  {Hafner}}, \bibinfo {author} {\bibfnamefont {P.}~\bibnamefont {Khanenko}},
  \bibinfo {author} {\bibfnamefont {E.-O.}\ \bibnamefont {Eljaouhari}},
  \bibinfo {author} {\bibfnamefont {R.}~\bibnamefont {Küchler}}, \bibinfo
  {author} {\bibfnamefont {J.}~\bibnamefont {Banda}}, \bibinfo {author}
  {\bibfnamefont {N.}~\bibnamefont {Bannor}}, \bibinfo {author} {\bibfnamefont
  {T.}~\bibnamefont {Lühmann}}, \bibinfo {author} {\bibfnamefont
  {J.}~\bibnamefont {Landaeta}}, \bibinfo {author} {\bibfnamefont
  {S.}~\bibnamefont {Mishra}}, \bibinfo {author} {\bibfnamefont
  {I.}~\bibnamefont {Sheikin}}, \bibinfo {author} {\bibfnamefont
  {E.}~\bibnamefont {Hassinger}}, \bibinfo {author} {\bibfnamefont
  {S.}~\bibnamefont {Khim}}, \bibinfo {author} {\bibfnamefont {C.}~\bibnamefont
  {Geibel}}, \bibinfo {author} {\bibfnamefont {G.}~\bibnamefont {Zwicknagl}},\
  and\ \bibinfo {author} {\bibfnamefont {M.}~\bibnamefont {Brando}},\ }\href
  {https://doi.org/10.1103/PhysRevX.12.011023} {\bibfield  {journal} {\bibinfo
  {journal} {Physical Review X}\ }\textbf {\bibinfo {volume} {12}},\ \bibinfo
  {pages} {011023} (\bibinfo {year} {2022})}\BibitemShut {NoStop}%
\bibitem [{\citenamefont {Kibune}\ \emph {et~al.}(2022)\citenamefont {Kibune},
  \citenamefont {Kitagawa}, \citenamefont {Kinjo}, \citenamefont {Ogata},
  \citenamefont {Manago}, \citenamefont {Taniguchi}, \citenamefont {Ishida},
  \citenamefont {Brando}, \citenamefont {Hassinger}, \citenamefont {Rosner},
  \citenamefont {Geibel},\ and\ \citenamefont {Khim}}]{kibune22}%
  \BibitemOpen
  \bibfield  {author} {\bibinfo {author} {\bibfnamefont {M.}~\bibnamefont
  {Kibune}}, \bibinfo {author} {\bibfnamefont {S.}~\bibnamefont {Kitagawa}},
  \bibinfo {author} {\bibfnamefont {K.}~\bibnamefont {Kinjo}}, \bibinfo
  {author} {\bibfnamefont {S.}~\bibnamefont {Ogata}}, \bibinfo {author}
  {\bibfnamefont {M.}~\bibnamefont {Manago}}, \bibinfo {author} {\bibfnamefont
  {T.}~\bibnamefont {Taniguchi}}, \bibinfo {author} {\bibfnamefont
  {K.}~\bibnamefont {Ishida}}, \bibinfo {author} {\bibfnamefont
  {M.}~\bibnamefont {Brando}}, \bibinfo {author} {\bibfnamefont
  {E.}~\bibnamefont {Hassinger}}, \bibinfo {author} {\bibfnamefont
  {H.}~\bibnamefont {Rosner}}, \bibinfo {author} {\bibfnamefont
  {C.}~\bibnamefont {Geibel}},\ and\ \bibinfo {author} {\bibfnamefont
  {S.}~\bibnamefont {Khim}},\ }\href
  {https://doi.org/10.1103/PhysRevLett.128.057002} {\bibfield  {journal}
  {\bibinfo  {journal} {Physical Review Letters}\ }\textbf {\bibinfo {volume}
  {128}},\ \bibinfo {pages} {057002} (\bibinfo {year} {2022})}\BibitemShut
  {NoStop}%
\bibitem [{\citenamefont {Hazra}\ and\ \citenamefont
  {Coleman}(2022)}]{hazra22}%
  \BibitemOpen
  \bibfield  {author} {\bibinfo {author} {\bibfnamefont {T.}~\bibnamefont
  {Hazra}}\ and\ \bibinfo {author} {\bibfnamefont {P.}~\bibnamefont
  {Coleman}},\ }\href {http://arxiv.org/abs/2205.13529} {\bibfield  {journal}
  {\bibinfo  {journal} {arXiv:2205.13529}\ } (\bibinfo {year}
  {2022})}\BibitemShut {NoStop}%
\bibitem [{\citenamefont {Sheikin}\ \emph {et~al.}(2001)\citenamefont
  {Sheikin}, \citenamefont {Huxley}, \citenamefont {Braithwaite}, \citenamefont
  {Brison}, \citenamefont {Watanabe}, \citenamefont {Miyake},\ and\
  \citenamefont {Flouquet}}]{sheikin01}%
  \BibitemOpen
  \bibfield  {author} {\bibinfo {author} {\bibfnamefont {I.}~\bibnamefont
  {Sheikin}}, \bibinfo {author} {\bibfnamefont {A.}~\bibnamefont {Huxley}},
  \bibinfo {author} {\bibfnamefont {D.}~\bibnamefont {Braithwaite}}, \bibinfo
  {author} {\bibfnamefont {J.}~\bibnamefont {Brison}}, \bibinfo {author}
  {\bibfnamefont {S.}~\bibnamefont {Watanabe}}, \bibinfo {author}
  {\bibfnamefont {K.}~\bibnamefont {Miyake}},\ and\ \bibinfo {author}
  {\bibfnamefont {J.}~\bibnamefont {Flouquet}},\ }\href@noop {} {\bibfield
  {journal} {\bibinfo  {journal} {Phys. Rev. B}\ }\textbf {\bibinfo {volume}
  {64}},\ \bibinfo {pages} {220503R} (\bibinfo {year} {2001})}\BibitemShut
  {NoStop}%
\bibitem [{\citenamefont {Miclea}\ \emph {et~al.}(2006)\citenamefont {Miclea},
  \citenamefont {Nicklas}, \citenamefont {Parker}, \citenamefont {Maki},
  \citenamefont {Sarrao}, \citenamefont {Thompson}, \citenamefont {Sparn},\
  and\ \citenamefont {Steglich}}]{miclea06}%
  \BibitemOpen
  \bibfield  {author} {\bibinfo {author} {\bibfnamefont {C.~F.}\ \bibnamefont
  {Miclea}}, \bibinfo {author} {\bibfnamefont {M.}~\bibnamefont {Nicklas}},
  \bibinfo {author} {\bibfnamefont {D.}~\bibnamefont {Parker}}, \bibinfo
  {author} {\bibfnamefont {K.}~\bibnamefont {Maki}}, \bibinfo {author}
  {\bibfnamefont {J.~L.}\ \bibnamefont {Sarrao}}, \bibinfo {author}
  {\bibfnamefont {J.~D.}\ \bibnamefont {Thompson}}, \bibinfo {author}
  {\bibfnamefont {G.}~\bibnamefont {Sparn}},\ and\ \bibinfo {author}
  {\bibfnamefont {F.}~\bibnamefont {Steglich}},\ }\href
  {https://link.aps.org/doi/10.1103/PhysRevLett.96.117001} {\bibfield
  {journal} {\bibinfo  {journal} {Phys. Rev. Lett.}\ }\textbf {\bibinfo
  {volume} {96}},\ \bibinfo {pages} {117001} (\bibinfo {year}
  {2006})}\BibitemShut {NoStop}%
\bibitem [{\citenamefont {Kittaka}\ \emph {et~al.}(2016)\citenamefont
  {Kittaka}, \citenamefont {Aoki}, \citenamefont {Shimura}, \citenamefont
  {Sakakibara}, \citenamefont {Seiro}, \citenamefont {Geibel}, \citenamefont
  {Steglich}, \citenamefont {Tsutsumi}, \citenamefont {Ikeda},\ and\
  \citenamefont {Machida}}]{kittaka16}%
  \BibitemOpen
  \bibfield  {author} {\bibinfo {author} {\bibfnamefont {S.}~\bibnamefont
  {Kittaka}}, \bibinfo {author} {\bibfnamefont {Y.}~\bibnamefont {Aoki}},
  \bibinfo {author} {\bibfnamefont {Y.}~\bibnamefont {Shimura}}, \bibinfo
  {author} {\bibfnamefont {T.}~\bibnamefont {Sakakibara}}, \bibinfo {author}
  {\bibfnamefont {S.}~\bibnamefont {Seiro}}, \bibinfo {author} {\bibfnamefont
  {C.}~\bibnamefont {Geibel}}, \bibinfo {author} {\bibfnamefont
  {F.}~\bibnamefont {Steglich}}, \bibinfo {author} {\bibfnamefont
  {Y.}~\bibnamefont {Tsutsumi}}, \bibinfo {author} {\bibfnamefont
  {H.}~\bibnamefont {Ikeda}},\ and\ \bibinfo {author} {\bibfnamefont
  {K.}~\bibnamefont {Machida}},\ }\href
  {https://link.aps.org/doi/10.1103/PhysRevB.94.054514} {\bibfield  {journal}
  {\bibinfo  {journal} {Phys. Rev. B}\ }\textbf {\bibinfo {volume} {94}},\
  \bibinfo {pages} {054514} (\bibinfo {year} {2016})}\BibitemShut {NoStop}%
\bibitem [{\citenamefont {Chen}\ \emph {et~al.}(1984)\citenamefont {Chen},
  \citenamefont {Lambert}, \citenamefont {Maple}, \citenamefont {Fisk},
  \citenamefont {Smith}, \citenamefont {Stewart},\ and\ \citenamefont
  {Willis}}]{chen84}%
  \BibitemOpen
  \bibfield  {author} {\bibinfo {author} {\bibfnamefont {J.~W.}\ \bibnamefont
  {Chen}}, \bibinfo {author} {\bibfnamefont {S.~E.}\ \bibnamefont {Lambert}},
  \bibinfo {author} {\bibfnamefont {M.~B.}\ \bibnamefont {Maple}}, \bibinfo
  {author} {\bibfnamefont {Z.}~\bibnamefont {Fisk}}, \bibinfo {author}
  {\bibfnamefont {J.~L.}\ \bibnamefont {Smith}}, \bibinfo {author}
  {\bibfnamefont {G.~R.}\ \bibnamefont {Stewart}},\ and\ \bibinfo {author}
  {\bibfnamefont {J.~O.}\ \bibnamefont {Willis}},\ }\href
  {https://link.aps.org/doi/10.1103/PhysRevB.30.1583} {\bibfield  {journal}
  {\bibinfo  {journal} {Phys. Rev. B}\ }\textbf {\bibinfo {volume} {30}},\
  \bibinfo {pages} {1583} (\bibinfo {year} {1984})}\BibitemShut {NoStop}%
\bibitem [{\citenamefont {Thomas}\ \emph {et~al.}(1996)\citenamefont {Thomas},
  \citenamefont {Wand}, \citenamefont {L{\"u}hmann}, \citenamefont {Gegenwart},
  \citenamefont {Stewart}, \citenamefont {Steglich}, \citenamefont {Brison},
  \citenamefont {Buzdin}, \citenamefont {Gl{\'e}mot},\ and\ \citenamefont
  {Flouquet}}]{thomas96}%
  \BibitemOpen
  \bibfield  {author} {\bibinfo {author} {\bibfnamefont {F.}~\bibnamefont
  {Thomas}}, \bibinfo {author} {\bibfnamefont {B.}~\bibnamefont {Wand}},
  \bibinfo {author} {\bibfnamefont {T.}~\bibnamefont {L{\"u}hmann}}, \bibinfo
  {author} {\bibfnamefont {P.}~\bibnamefont {Gegenwart}}, \bibinfo {author}
  {\bibfnamefont {G.~R.}\ \bibnamefont {Stewart}}, \bibinfo {author}
  {\bibfnamefont {F.}~\bibnamefont {Steglich}}, \bibinfo {author}
  {\bibfnamefont {J.~P.}\ \bibnamefont {Brison}}, \bibinfo {author}
  {\bibfnamefont {A.}~\bibnamefont {Buzdin}}, \bibinfo {author} {\bibfnamefont
  {L.}~\bibnamefont {Gl{\'e}mot}},\ and\ \bibinfo {author} {\bibfnamefont
  {J.}~\bibnamefont {Flouquet}},\ }\href@noop {} {\bibfield  {journal}
  {\bibinfo  {journal} {J. Low Temp. Phys.}\ }\textbf {\bibinfo {volume}
  {102}},\ \bibinfo {pages} {117} (\bibinfo {year} {1996})}\BibitemShut
  {NoStop}%
\bibitem [{\citenamefont {Clogston}\ \emph {et~al.}(1962)\citenamefont
  {Clogston}, \citenamefont {Gossard}, \citenamefont {Jaccarino},\ and\
  \citenamefont {Yafet}}]{clogston62}%
  \BibitemOpen
  \bibfield  {author} {\bibinfo {author} {\bibfnamefont {A.~M.}\ \bibnamefont
  {Clogston}}, \bibinfo {author} {\bibfnamefont {A.~C.}\ \bibnamefont
  {Gossard}}, \bibinfo {author} {\bibfnamefont {V.}~\bibnamefont {Jaccarino}},\
  and\ \bibinfo {author} {\bibfnamefont {Y.}~\bibnamefont {Yafet}},\ }\href
  {https://link.aps.org/doi/10.1103/PhysRevLett.9.262} {\bibfield  {journal}
  {\bibinfo  {journal} {Phys. Rev. Lett.}\ }\textbf {\bibinfo {volume} {9}},\
  \bibinfo {pages} {262} (\bibinfo {year} {1962})}\BibitemShut {NoStop}%
\bibitem [{\citenamefont {Schossmann}\ and\ \citenamefont
  {Carbotte}(1989)}]{schossmann89}%
  \BibitemOpen
  \bibfield  {author} {\bibinfo {author} {\bibfnamefont {M.}~\bibnamefont
  {Schossmann}}\ and\ \bibinfo {author} {\bibfnamefont {J.~P.}\ \bibnamefont
  {Carbotte}},\ }\href {https://link.aps.org/doi/10.1103/PhysRevB.39.4210}
  {\bibfield  {journal} {\bibinfo  {journal} {Phys. Rev. B}\ }\textbf {\bibinfo
  {volume} {39}},\ \bibinfo {pages} {4210} (\bibinfo {year}
  {1989})}\BibitemShut {NoStop}%
\bibitem [{\citenamefont {Carbotte}(1990)}]{carbotte90}%
  \BibitemOpen
  \bibfield  {author} {\bibinfo {author} {\bibfnamefont {J.~P.}\ \bibnamefont
  {Carbotte}},\ }\href {https://link.aps.org/doi/10.1103/RevModPhys.62.1027}
  {\bibfield  {journal} {\bibinfo  {journal} {Rev. Mod. Phys.}\ }\textbf
  {\bibinfo {volume} {62}},\ \bibinfo {pages} {1027} (\bibinfo {year}
  {1990})}\BibitemShut {NoStop}%
\bibitem [{\citenamefont {P\'erez-Gonz\'alez}(1996)}]{perez-gonzalez96}%
  \BibitemOpen
  \bibfield  {author} {\bibinfo {author} {\bibfnamefont {A.}~\bibnamefont
  {P\'erez-Gonz\'alez}},\ }\href {https://doi.org/10.1103/PhysRevB.54.16053}
  {\bibfield  {journal} {\bibinfo  {journal} {Physical Review B}\ }\textbf
  {\bibinfo {volume} {54}},\ \bibinfo {pages} {16053} (\bibinfo {year}
  {1996})}\BibitemShut {NoStop}%
\bibitem [{\citenamefont {Schossmann}\ and\ \citenamefont
  {Schachinger}(1986)}]{schossmann86}%
  \BibitemOpen
  \bibfield  {author} {\bibinfo {author} {\bibfnamefont {M.}~\bibnamefont
  {Schossmann}}\ and\ \bibinfo {author} {\bibfnamefont {E.}~\bibnamefont
  {Schachinger}},\ }\href {https://link.aps.org/doi/10.1103/PhysRevB.33.6123}
  {\bibfield  {journal} {\bibinfo  {journal} {Phys. Rev. B}\ }\textbf {\bibinfo
  {volume} {33}},\ \bibinfo {pages} {6123} (\bibinfo {year}
  {1986})}\BibitemShut {NoStop}%
\bibitem [{\citenamefont {Matsumoto}\ \emph {et~al.}(2012)\citenamefont
  {Matsumoto}, \citenamefont {Koga},\ and\ \citenamefont
  {Kusunose}}]{matsumoto12}%
  \BibitemOpen
  \bibfield  {author} {\bibinfo {author} {\bibfnamefont {M.}~\bibnamefont
  {Matsumoto}}, \bibinfo {author} {\bibfnamefont {M.}~\bibnamefont {Koga}},\
  and\ \bibinfo {author} {\bibfnamefont {H.}~\bibnamefont {Kusunose}},\ }\href
  {http://journals.jps.jp/doi/10.1143/JPSJ.81.033702} {\bibfield  {journal}
  {\bibinfo  {journal} {J. Phys. Soc. Jpn.}\ }\textbf {\bibinfo {volume}
  {81}},\ \bibinfo {pages} {033702} (\bibinfo {year} {2012})}\BibitemShut
  {NoStop}%
\bibitem [{\citenamefont {Aperis}\ \emph {et~al.}(2015)\citenamefont {Aperis},
  \citenamefont {Maldonado},\ and\ \citenamefont {Oppeneer}}]{aperis15}%
  \BibitemOpen
  \bibfield  {author} {\bibinfo {author} {\bibfnamefont {A.}~\bibnamefont
  {Aperis}}, \bibinfo {author} {\bibfnamefont {P.}~\bibnamefont {Maldonado}},\
  and\ \bibinfo {author} {\bibfnamefont {P.~M.}\ \bibnamefont {Oppeneer}},\
  }\href {https://link.aps.org/doi/10.1103/PhysRevB.92.054516} {\bibfield
  {journal} {\bibinfo  {journal} {Phys. Rev. B}\ }\textbf {\bibinfo {volume}
  {92}},\ \bibinfo {pages} {054516} (\bibinfo {year} {2015})}\BibitemShut
  {NoStop}%
\bibitem [{\citenamefont {Aperis}\ \emph {et~al.}(2020)\citenamefont {Aperis},
  \citenamefont {Morooka},\ and\ \citenamefont {Oppeneer}}]{aperis20}%
  \BibitemOpen
  \bibfield  {author} {\bibinfo {author} {\bibfnamefont {A.}~\bibnamefont
  {Aperis}}, \bibinfo {author} {\bibfnamefont {E.~V.}\ \bibnamefont
  {Morooka}},\ and\ \bibinfo {author} {\bibfnamefont {P.~M.}\ \bibnamefont
  {Oppeneer}},\ }\href
  {https://linkinghub.elsevier.com/retrieve/pii/S0003491620300282} {\bibfield
  {journal} {\bibinfo  {journal} {Annals of Physics}\ }\textbf {\bibinfo
  {volume} {417}},\ \bibinfo {pages} {168095} (\bibinfo {year}
  {2020})}\BibitemShut {NoStop}%
\bibitem [{\citenamefont {Helfand}\ and\ \citenamefont
  {Werthamer}(1966)}]{helfand66}%
  \BibitemOpen
  \bibfield  {author} {\bibinfo {author} {\bibfnamefont {E.}~\bibnamefont
  {Helfand}}\ and\ \bibinfo {author} {\bibfnamefont {N.~R.}\ \bibnamefont
  {Werthamer}},\ }\href {https://link.aps.org/doi/10.1103/PhysRev.147.288}
  {\bibfield  {journal} {\bibinfo  {journal} {Phys. Rev.}\ }\textbf {\bibinfo
  {volume} {147}},\ \bibinfo {pages} {288} (\bibinfo {year}
  {1966})}\BibitemShut {NoStop}%
\bibitem [{\citenamefont {Tinkham}(2004)}]{tinkham04}%
  \BibitemOpen
  \bibfield  {author} {\bibinfo {author} {\bibfnamefont {M.}~\bibnamefont
  {Tinkham}},\ }\href@noop {} {\emph {\bibinfo {title} {Introduction to
  Superconductivity}}}\ (\bibinfo  {publisher} {{Dover Publications}},\
  \bibinfo {year} {2004})\BibitemShut {NoStop}%
\bibitem [{\citenamefont {Orlando}\ \emph {et~al.}(1979)\citenamefont
  {Orlando}, \citenamefont {McNiff}, \citenamefont {Foner},\ and\ \citenamefont
  {Beasley}}]{orlando79}%
  \BibitemOpen
  \bibfield  {author} {\bibinfo {author} {\bibfnamefont {T.~P.}\ \bibnamefont
  {Orlando}}, \bibinfo {author} {\bibfnamefont {E.~J.}\ \bibnamefont {McNiff}},
  \bibinfo {author} {\bibfnamefont {S.}~\bibnamefont {Foner}},\ and\ \bibinfo
  {author} {\bibfnamefont {M.~R.}\ \bibnamefont {Beasley}},\ }\href
  {https://link.aps.org/doi/10.1103/PhysRevB.19.4545} {\bibfield  {journal}
  {\bibinfo  {journal} {Phys. Rev. B}\ }\textbf {\bibinfo {volume} {19}},\
  \bibinfo {pages} {4545} (\bibinfo {year} {1979})}\BibitemShut {NoStop}%
\bibitem [{\citenamefont {Sullivan}\ and\ \citenamefont
  {Seidel}(1968)}]{sullivan68}%
  \BibitemOpen
  \bibfield  {author} {\bibinfo {author} {\bibfnamefont {P.~F.}\ \bibnamefont
  {Sullivan}}\ and\ \bibinfo {author} {\bibfnamefont {G.}~\bibnamefont
  {Seidel}},\ }\href@noop {} {\bibfield  {journal} {\bibinfo  {journal}
  {Physical Review}\ }\textbf {\bibinfo {volume} {173}},\ \bibinfo {pages}
  {679} (\bibinfo {year} {1968})}\BibitemShut {NoStop}%
\bibitem [{\citenamefont {Gati}\ \emph {et~al.}(2019)\citenamefont {Gati},
  \citenamefont {Drachuck}, \citenamefont {Xiang}, \citenamefont {Wang},
  \citenamefont {Bud'ko},\ and\ \citenamefont {Canfield}}]{gati19}%
  \BibitemOpen
  \bibfield  {author} {\bibinfo {author} {\bibfnamefont {E.}~\bibnamefont
  {Gati}}, \bibinfo {author} {\bibfnamefont {G.}~\bibnamefont {Drachuck}},
  \bibinfo {author} {\bibfnamefont {L.}~\bibnamefont {Xiang}}, \bibinfo
  {author} {\bibfnamefont {L.-L.}\ \bibnamefont {Wang}}, \bibinfo {author}
  {\bibfnamefont {S.~L.}\ \bibnamefont {Bud'ko}},\ and\ \bibinfo {author}
  {\bibfnamefont {P.~C.}\ \bibnamefont {Canfield}},\ }\href
  {http://aip.scitation.org/doi/10.1063/1.5084730} {\bibfield  {journal}
  {\bibinfo  {journal} {Review of Scientific Instruments}\ }\textbf {\bibinfo
  {volume} {90}},\ \bibinfo {pages} {023911} (\bibinfo {year}
  {2019})}\BibitemShut {NoStop}%
\end{thebibliography}%



\end{document}